\documentclass[aps,prl,twocolumn,showpacs,amsmath,amssymb]{revtex4}

\usepackage{graphicx}
\usepackage{dcolumn}
\usepackage{bm}
\usepackage{textcomp,mathcomp}

\begin{document}

\title{Macroscopic Entanglement of a Bose Einstein Condensate on a Superconducting Atom Chip}

\author{Mandip Singh}
\email{masingh@swin.edu.au}

\affiliation{
\\Centre for Atom Optics and Ultrafast Spectroscopy and
\\ ARC Centre of Excellence for Quantum-Atom Optics
\\ Swinburne University of Technology, Hawthorn, Victoria 3122, Australia }

\date{\today}

\begin{abstract}

We propose and analyse a practically implementable scheme to generate macroscopic entanglement of a Bose-Einstein condensate in a micro-magnetic trap magnetically coupled to a superconducting loop. We treat the superconducting loop in a quantum superposition of two different flux states coupled with the magnetic trap to generate macroscopic entanglement. Our scheme also provides a platform to realise interferometry of entangled atoms through the Bose-Einstein condensate and to explore physics at the quantum-classical interface.

\end{abstract}

\pacs{03.65.Ud, 03.75.Gg,03.67.-a,03.75.Dg,39.20.+q,03.75.Nt }

\maketitle


Entanglement is considered to be one of the most fundamental features of quantum mechanics. In addition, it is of great importance in the context of quantum information and quantum computation. In recent years, there have been considerable efforts to generate and preserve entanglement for quantum information processing. In particular, entanglement at macroscopic level is of prime interest to explore physics at the interface of classical and quantum mechanics. Also, macroscopically entangled states are promising candidates for the practical realization of a quantum computer.

In recent years there has been ground breaking progress in the field of manipulation of BECs. Nowadays, it is relatively easy to produce a BEC in a micro-magnetic trap on an atom chip \cite{han01}. A BEC in such traps can be coherently manipulated with RF fields \cite{sch05, hoff06} and microwaves \cite{tre06, tre04}. Since neutral atoms can be positioned a few microns from the chip surface and moved with nanometre resolution, atom chips provide a convenient platform to study the interaction between a BEC and a nearby surface \cite{jon03} including the study of fundamental quantum effects such as the Casimir-Polder interaction \cite{har05}. The field of superconducting circuits is also progressing rapidly in terms of technological implementation and realization of quantum coherent control of superconducting qubits \cite{mak01, moo99}. A macroscopic superposition of different magnetic flux states has been demonstrated \cite{fri00, van00} and quantum coherent dynamics of flux qubits have been realised \cite{chi03}. Entanglement and decoherence of a micromechanical resonator via coupling to a Cooper-pair box have been studied \cite{arm02}. Recently, a mechanical analogue of cavity quantum electrodynamics was proposed based on the idea of coupling a nanomechanical resonator to the BEC on an atom chip \cite{tre07}. Atom chips based on superconducting substrates and wires have been implemented \cite{nirr06, muk07}. However, to our knowledge there has been no proposal or realization of quantum coherent dynamics based on the quantum mechanical properties of a superconducting circuit coupled with the ultracold atoms on an atom chip. Thus, by exploiting the quantum mechanical properties of superconducting circuits and their interactions with nearby ultracold atoms or a BEC new physics at the quantum-classical interface can be explored.

\begin{figure}
\includegraphics[scale=0.3]{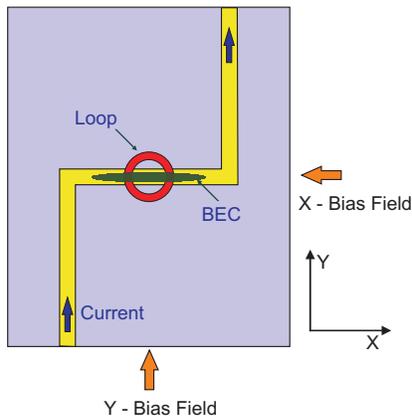}
\caption{\label{fig:chip}(Color online) Schematic showing a superconducting loop situated symmetrically above a Z-shaped wire magnetic trap. The magnetic trap containing a BEC can be moved closer to the loop in order to increase the interaction between them. The superconducting loop can be biased appropriately by applying an external field along the z-direction (pointing normal to the page).}
\end{figure}

In this Letter we propose a scheme which combines two emerging fields, micro-manipulation of a BEC on an atom chip and quantum coherent control in superconducting circuits, to realise a macroscopic entanglement of a BEC. Our proposal is based on coupling a superconducting loop to a magnetic trap containing a BEC on an atom chip. The physical arrangement is shown in Fig.~\ref{fig:chip}. Ultracold neutral atoms or a BEC can be trapped in a magnetic trap above the surface by applying an external bias field in the plane of a Z-shaped current carrying wire. A superconducting loop is positioned symmetrically above the Z-wire and below the trap in such a way that there is zero net flux linked to it. The trap position and trap frequencies depend on the external bias fields and the current in the trapping wire. A nearby superconducting loop carrying persistent current can perturb such a magnetic trap through magnetostatic interactions. The sign of such a perturbation depends on the direction of the persistent current flowing in the loop and its location with respect to the trap.

A superconducting loop when placed in an external flux permits only discrete values of the net flux threaded through it, which is an integral multiple of the flux quantum. In other words, the super-current in the loop responds automatically to any change in the externally applied flux in order to keep the closed loop phase acquired by the wave function an integral multiple of 2$\pi$ . In the case of a superconducting ring interrupted by a Josephson junction the total energy corresponds to a double well in the flux basis \cite{fri00}, where the left (right) well corresponds to persistent current flowing in a clockwise (anticlockwise) direction in the loop. When the external flux is equal to half of the flux quantum such a double well is symmetric. The inter-well tunneling can be controlled by replacing the Josephson junction with split junctions (DC-SQUID) \cite{fri00}.

At low temperature the superconducting  loop can be prepared in a quantum superposition of two persistent current states flowing clockwise and anticlockwise by biasing it at half of the flux quantum. If the magnetostatic coupling of the loop to a magnetic trap containing a BEC is increased adiabatically, the BEC will follow the trap perturbation caused by the persistent current in the loop. Since the persistent current is in quantum superposition, a macroscopic entanglement between the state of the BEC in the perturbed trap configurations and the persistent current state of the superconducting loop can be created. One trap configuration can differ from the other one in terms of its spatial distribution and the chemical potential of the BEC. The entanglement can be detected by carefully measuring the centre of mass distribution of the released condensate in repeated measurements. The distribution of the centre of mass is a consequence of interference of the entangled atoms, which is also discussed in this Letter.

The entangling dynamics can be described by the Hamiltonian of the superconducting loop coupled to the magnetic trap containing a BEC. In the symmetric case the Hamiltonian of the superconducting loop can be treated as a two level system at low temperature
\begin{equation}
\label{eq:1} H_{S}=E_{0}|0\rangle\langle0|+E_{0}|1\rangle\langle1|+J|1\rangle\langle0|+J|0\rangle\langle1|
\end{equation}
where $|0\rangle$ and $|1\rangle$ represent the ground state of the left and the right well, respectively and $J$ is the tunnelling amplitude between them.

The Hamiltonian of an atom of mass $m$ in the ground state of the trap coupled to the superconducting loop in the case $J = 0$ can be written as

\begin{eqnarray}
\label{eq:2}
 H_{T}&=& \int\hat{\Psi}^{\dag}(r){\Bigg[}\left(\frac{-\hbar^2}{2m}\nabla^2+ V(r)\right)\hat{I}+\Delta V_{0}(r)|0\rangle\langle0| \nonumber \\[3mm]
      & & +\ \Delta V_{1}(r)|1\rangle\langle1| {\Bigg]}\hat{\Psi}(r) \mathrm{d}r
\end{eqnarray}

where $\Delta V_{0}(r)$ and $\Delta V_{1}(r)$ are the perturbations in the trap potential $\Delta V(r)$ due to interaction with the loop for states   and $|0\rangle$, $|1\rangle$ respectively and $|0\rangle \langle0| + |1\rangle \langle1| = \hat{I} $. In the case when the perturbation is increased adiabatically the field operator $\hat{\Psi}(r,t)$ can be expanded as a linear combination of $\phi_{0}(r,t)$  and $\phi_{1}(r,t)$ the ground states in the two perturbed potentials, respectively;

\begin{equation}
\label{eq:3} \hat{\Psi}(r,t)=\hat{a}_{0}\phi_{0}(r,t)|0\rangle\langle0|+\hat{a}_{1}\phi_{1}(r,t)|1\rangle\langle1|
\end{equation}

where $\hat{a}_{0}$ and $\hat{a}_{1}$   are the corresponding bosonic annihilation operators.
Therefore, from Eq.~\ref{eq:2}, Eq.~\ref{eq:3} and Eq.~\ref{eq:4} the total Hamiltonian of the system can be written as

\begin{equation}
\label{eq:4} H=E_{0}|0\rangle\langle0|+E_{0}|1\rangle\langle1|+\mu_{0}(t)\hat{a}^\dag_{0}\hat{a_{0}}|0\rangle\langle0|+ \mu_{1}(t)\hat{a}^{\dag}_{1}\hat{a}_{1}|1\rangle\langle1|
\end{equation}

where

\begin{equation}
\label{eq:5} \mu_{0}(t)=\int\phi^\dag_{0}(r,t)\left(\frac{-\hbar^2}{2m}\nabla^2+V(r)+\Delta V_{0}(r,t)\right)\phi_{0}(r,t) \mathrm{d}r
\end{equation}

and

\begin{equation}
\label{eq:6} \mu_{1}(t)=\int\phi^\dag_{1}(r,t)\left(\frac{-\hbar^2}{2m}\nabla^2+V(r)+\Delta V_{1}(r,t)\right)\phi_{1}(r,t) \mathrm{d}r
\end{equation}

are the energy eigenvalues in the case of two perturbed situations of the trap.

            Let us see how entanglement can be generated. In the first step a BEC of $N$ atoms can be prepared in a Z-wire magnetic trap far away from the superconducting loop so that coupling between them can be neglected. In the second step the superconducting loop can be prepared in a symmetric superposition $|S\rangle=(|0\rangle + |1\rangle)/\sqrt{2}$  and then the tunnelling amplitude $J$  can be reduced to zero. In the third step the BEC can be slowly brought closer to the superconducting loop so that the coupling between them is increased adiabatically and BEC will follow two different configurations of the trap in quantum superposition. At one point where the coupling is sufficiently strong they will be distinguishable.

\begin{figure}
\includegraphics[scale=0.3]{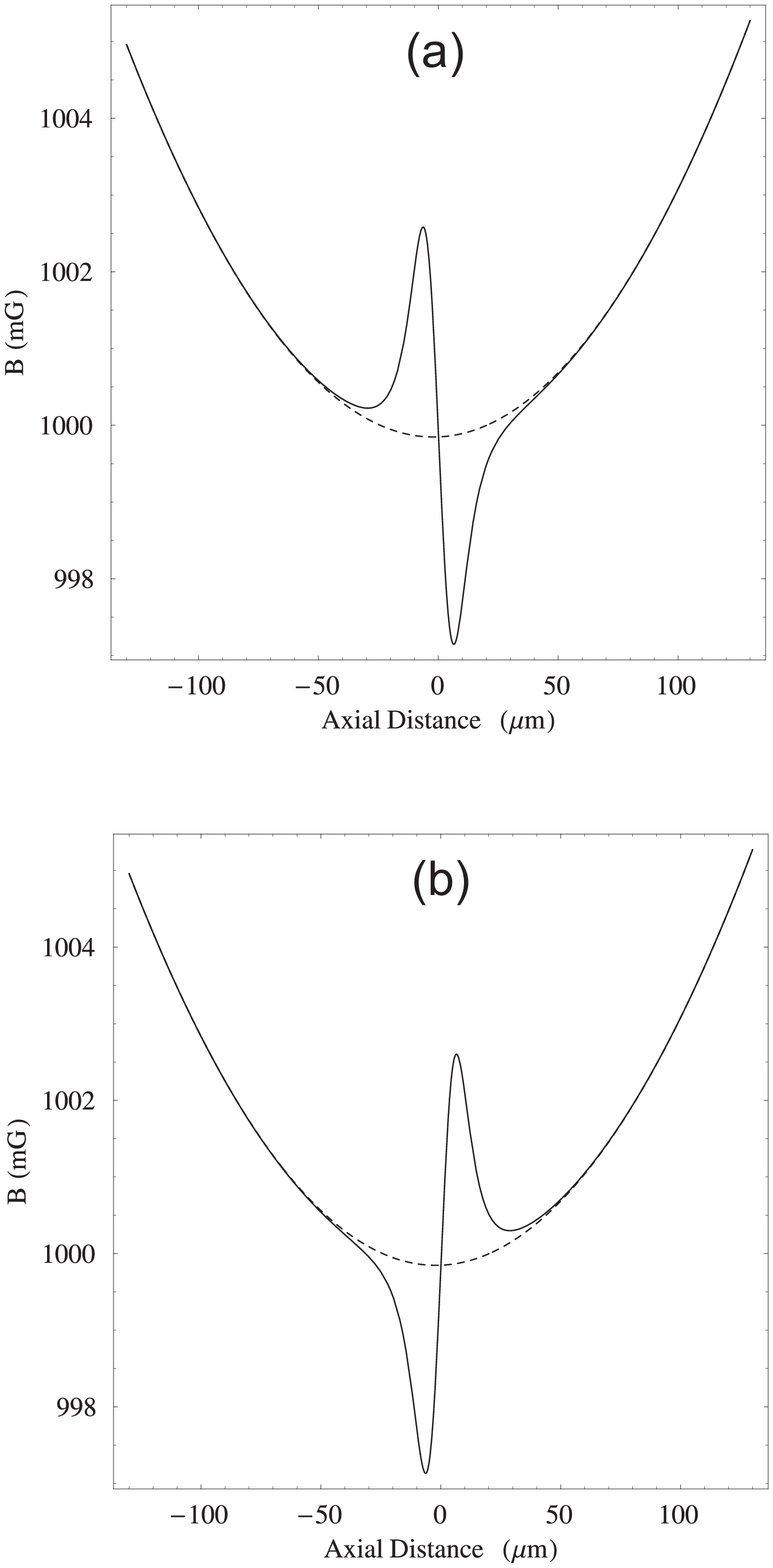}
\caption{\label{fig:clock}Magnetic field profile of a Z-wire magnetic trap along the axial direction coupled to a superconducting loop when the persistent current flows (a) clockwise and (b) anti-clockwise. The dotted curve represents the field profile without any coupling.}
\end{figure}

Therefore at $t=0$ the initial state of the system is $|\Psi,t=0\rangle=(|0\rangle+|1\rangle)|N,t=0\rangle/\sqrt{2}$,
where $|N,t=0\rangle$   is the state corresponding to $N$ atoms in the ground state of the trap in the case of no coupling with the flux loop. When the coupling is increased adiabatically the state of the system evolves to

\begin{eqnarray}
\label{eq:7} |\Psi,t\rangle & =& \frac{e^{i\gamma_{0}(t)-i(E_{0}t+N\int_0^{t}\mu_{0}(t') \mathrm{d}t')/\hbar}}{\sqrt{2}}\nonumber \\[3mm]
& &    \times \left[|0\rangle|N,t\rangle_{0}+e^{i\Phi(t)}|1\rangle|N,t\rangle_{1}\right]
\end{eqnarray}

where

\begin{equation}
\label{eq:8} \Phi(t)=N    \int_0^{t}\frac{\mu_{0}(t')-\mu_{1}(t')}{\hbar} \mathrm{d}t'+\gamma_{1}(t)-\gamma_{0}(t)
\end{equation}

   also $\gamma_{0} (t)$ and $\gamma_{1}(t)$ represent the geometrical phase and $|N,t\rangle_{0} $ and $|N,t\rangle_{1} $ are states corresponding to $N$ atoms in the BEC in two perturbed situations of the trap, respectively. Since the process is adiabatic these two states follow the ground state of their respective perturbed potentials with time and when the coupling is zero these two states correspond to $|N,t\rangle $ . It is evident that state $|\Psi,t\rangle $  spans different macroscopically entangled states as $\Phi(t)$ evolves with time. If we only consider the state of the condensate for $\Phi(t)=0 $ and ignore the overall phase, the state (4) can also be expressed as

   \begin{equation}
\label{eq:9} |\Psi\rangle=\frac{1}{\sqrt{2}}\left(|0,N\rangle_{01}+|N,0\rangle_{01}\right)
\end{equation}

where $|0,N\rangle_{01} $ is the state representing zero atoms in one perturbed configuration of the trap and $N$ atoms in the other configuration. The perturbation in the Z-wire magnetic trap potential along the axial direction due to the superconducting loop is shown in Fig.~\ref{fig:clock}. The distance between the superconducting loop and the trap centre, which is located on the axis of the loop, is $10 $~\textmu m and the diameter of the superconducting loop is $10$ ~\textmu m. A field of about 0.1 G applied along the z-direction to bias the loop. The trap parameters are calculated for a Z-wire of length 5 mm, 5 A current and 20 G bias field along the y-direction.  The trap bottom can be adjusted by applying a field in the x-direction. For these parameters the amplitude of the perturbation is about 5.5 mG and the chemical potential for $N$ atoms of $^{87}$Rb (for $|F = 2, m_{F} = 2\rangle$) in absence of any perturbation is $0.02631N^{2/5}$ mG. The distance between the minima of the two perturbed configurations is of the order of the diameter of the superconducting loop.

The next important question is how to detect such a state and how it can be distinguished from a non-entangled state. The quantum interference of such an entangled state has been demonstrated for photons \cite{wal04, mit04} and studied for ions \cite{mol99, sac00}. The quantum correlation properties of multi-particle systems are discussed in \cite{bach04}. By releasing a condensate existing in an entangled state  Eq.~\ref{eq:9} one can observe quantum interference between different base vectors. However, in contrast to the case of the interference pattern observed for two different condensates \cite{shin04} the interference from the state  Eq.~\ref{eq:9} has some characteristic properties. The interference pattern obtained after time $t$ by spatially overlapping two different condensates (with initial Gaussian width $\sigma_{0} $ ) located at $\pm d/2$  has a density modulation with the inverse of its periodicity given by $1/\Lambda=tmd/(2\pi\hbar(t^2+(m\sigma^2_{0}/\hbar)^2))$ . However, for entangled state  Eq.~\ref{eq:9} the interference can be treated as one large particle, which is made of $N$ particles, interfering with itself. In this case the periodicity in the centre of mass density pattern obtained by repeated measurements is  $\Lambda/N$ and the system behaves as if its de Broglie wavelength is diminished by a factor of $N$. This interference pattern is a signature of the existence of macroscopic entanglement of the BEC with the superconducting loop. However, only those measurements which count same $N$ should be considered, therefore, a precise control for state preparation (Eq.~\ref{eq:7}) is required, where $\Phi(t)$ should not vary more than over $\pi/(2N)$ from measurement to measurement.  Details of similar measurements are discussed in \cite{bach04}. However, the number of particles in state Eq.~\ref{eq:9} are limited by the requirement of higher resolution and better detection efficiency. This also demands improved shot to shot phase stability. Another parameter which could also limit the number of particles is the back action of the atoms on the superconducting loop which we have not considered in our treatment. On the other hand, an appropriate operation on the state of the superconducting loop can also provide an evidence of the existence of a macroscopic entanglement of the coupled system.

In the context of experimental realization it is important to consider that the magnetic field from the Z-wire and the bias field should be less than the critical field of the superconducting loop. The experiment can be constructed by utilizing flip-chip technology, where the Z-wire and the superconducting loop can be constructed on two different substrates which can be bonded together with high precision with an appropriate gap between them. The Z-wire can also be constructed from a superconducting material in order to reduce the technical noise in the current. The atom chip should be shielded from background radiation by a gold coated copper shield and the whole assembly can be mounted on a cold finger with the chip pointing up side down. The ultracold atoms can be prepared in a different chamber and magnetically transported to the chip where they can be trapped and evaporatevely cooled \cite{muk07} down to BEC. There are various factors which can destroy macroscopic entangled state described by Eq.~\ref{eq:9}. Loss of a single atom from this state can easily destroy it by sharing its information with the environment. However, a superconducting flux superposition itself is prone to environment induced decoherence \cite{mak01}. The process of entangling the BEC with the superconducting loop must happen within the decoherence time limit. In the case of a short decoherence time the adiabatic condition might be interrupted, therefore the BEC could be excited.

In conclusion we have shown how the coherent dynamics of a superconducting loop can be used to generate a macroscopic entanglement of a BEC on an atom chip. Such a macroscopic entanglement could be useful to explore fundamental quantum mechanics by studying how quantum mechanical effects behave at the macroscopic level and how they decohere with the size of the system. It may also be possible to explore decoherence between the superconducting circuit and the BEC by varying the trap parameters and the size of the system such as the number of atoms in the BEC. Also, after the generation of entanglement the BEC could be excited in order to study how randomness in a macroscopic system plays a role in the destruction of quantum coherence. Finally, we have shown how such a macroscopic entanglement can be detected by exploiting the interference pattern of an entangled BEC. Thus, our scheme also provides a way to realize quantum entanglement interferometry on an atom chip.

\begin{acknowledgments}
The author is very thankful to Prof Peter Hannaford and Prof Tien Kieu for useful and stimulating discussions.
\end{acknowledgments}

\bibliography{mandip_resub}

\end{document}